\documentclass[11pt,twocolumn]{article}
\usepackage{a4}
\usepackage{times}
\usepackage{mathptm}
\usepackage[dvips]{epsfig}
\usepackage[dvips]{rotating}
\usepackage{natbib}
\usepackage{times}
\usepackage{fancyheadings}
\usepackage{tabularx}
\bibpunct{(}{)}{;}{a}{}{,}

\newcommand{\he}{HE~0107$-$5240}
\newcommand{\cd}{CD~$-38^{\circ}\,245$}
\newcommand{\tefft}{$T_{\mbox{\scriptsize eff}}$}
\newcommand{\teffm}{T_{\mbox{\scriptsize eff}}}
\newlength{\captionbreite}
\begin{document}
\noindent\raisebox{2cm}[-2cm]{\emph{Nature} {\bf 419} (2002), 904--906 (issue 31 October)}
\begin{flushleft}
  {\bf \Large A stellar relic from}\\[0.5ex]
  {\bf \Large the early Milky Way}\\[1ex]
  N. Christlieb\footnotemark[1]\footnotemark[3], M.S. Bessell\footnotemark[4],
  T.C. Beers\footnotemark[5],\\ B. Gustafsson\footnotemark[3], A. Korn\footnotemark[6],
  P.S. Barklem\footnotemark[3],\\ T. Karlsson\footnotemark[3],
  M. Mizuno-Wiedner\footnotemark[3] \& S. Rossi\footnotemark[7]\\[1ex]
  
  \footnotetext[1]{Hamburger Sternwarte, Gojenbergsweg 112, D-21029 Hamburg,
    Germany}
  \footnotetext[3]{Department of Astronomy and Space Physics, Uppsala
    University, Box 524, SE-75120 Uppsala, Sweden}
  \footnotetext[4]{Research School of Astonomy and Astrophysics, Mount
    Stromlo Observatory, Cotter Road, Weston, ACT 2611, Australia}
  \footnotetext[5]{Department of Physics and Astronomy, Michigan State
    University, East Lansing, MI 48824, USA}
  \footnotetext[6]{Universit\"ats-Sternwarte M\"unchen, Scheinerstrasse 1,\\
    D-81679 M\"unchen, Germany}
  \footnotetext[7]{Instituto de Astronomia, Geof\'{\i}sica e Ci\c{e}ncias
    Atmosf\'ericas, Departamento de Astronomia, Universidade de S\~ao Paulo, 
    05508-900 S\~ao Paulo, Brazil }
\end{flushleft}

{\bf The chemical composition of the most metal-deficient stars reflects the
  composition of the gas from which they formed. These old stars provide
  crucial clues to the star formation history and the synthesis of chemical
  elements in the early Universe. They are the local relics of epochs
  otherwise observable only at very high redshifts$^{1,2}$; if totally
  metal-free (``population III'') stars could be found, this would
  allow the direct study of the pristine gas from the Big Bang. Earlier searches
  for such stars found none with an iron abundance less than 1/10,000
  that of the Sun$^{3,4}$, leading to the suggestion$^{5,6}$ that low-mass
  stars could only form from clouds above a critical iron abundance. 
  Here we report the discovery of a low-mass star with an
  iron abundance as low as 1/200,000 of the solar value. This discovery
  suggests that population III stars could still exist, that is, that the first
  generation of stars also contained long-lived low-mass objects. The previous
  failure to find them may be an observational selection effect.}

The star {\he}, at coordinates right ascension
$\mbox{R.A.}(2000.0)=01\,\mbox{h}\;09\,\mbox{m}\;29.1\,\mbox{s}$ and
declination $\delta=-52^{\circ}\;24'\;34''$, is a giant star of the Galactic
halo population with apparent magnitude $B=15.86$. It was found during
medium-resolution spectroscopic follow-up observations of candidate metal-poor
stars selected from the Hamburg/ESO objective prism survey (HES)$^{7,8}$. This
survey, which covers the entire southern high-galactic-latitude sky to an
apparent magnitude limit of $B\approx 17.5$, extends the total survey volume
for metal-poor stars in the Galaxy by almost one order of magnitude compared
to the total volume explored by previous spectroscopic surveys.

A medium-resolution ($\delta\lambda \approx 0.2$\,nm) spectrum of {\he} was
obtained by M.S.B. with the Siding Spring Observatory 2.3-m telescope on
12 November 2001. The Ca\,II~K ($\lambda =393.4$\,nm) line was barely visible
in that spectrum, indicating that the star was likely to be extremely metal
deficient. Shortly thereafter, a high-resolution, high signal-to-noise ratio
spectrum was obtained with the 8-m Unit Telescope 2 (UT2) of the Very Large
Telescope at the European Southern Observatory (ESO), Paranal, Chile.

\begin{figure}[htb]
  \epsfig{file=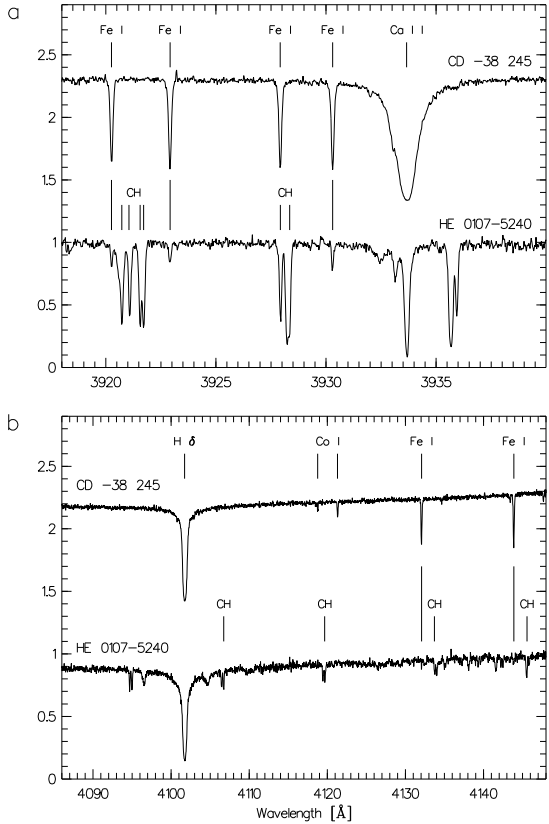, clip=, width=7.3cm,
    bbllx=170, bblly=288, bburx=427, bbury=677}
  \centering
  \caption{\footnotesize A portion of the spectrum of {\he}, shown compared to the
    spectrum of {\cd}, the previously most iron-poor giant star known.  Both
    spectra were obtained with VLT-UT2, and the Ultraviolet-Visual Echelle
    Spectrograph (UVES). We note the strong molecular CH and C$_2$ lines and
    extremely weak lines of Fe\,I in the spectrum of {\he}. The spectra used
    in our analysis have a resolution of $R = \lambda/\Delta\lambda = 40,000$,
    and a signal-to-noise ratio of more than 100 per pixel at $\lambda >
    400.0$\,nm. The covered wavelength ranges are $329.0$--$452.0$\,nm,
    $478.0$--$576.0$\,nm, and $583.0$--$681.0$\,nm. }
\end{figure}

We derive an effective temperature $\teffm = 5100\pm150$\,K for {\he} by means
of broad-band visual and infrared photometry. The absence of Fe\,II lines,
through the Fe\,I/Fe\,II ionisation equilibrium, constrains the star to have a
logarithmic surface gravity of $\log(g) > 2.0$\,dex, while main-sequence
gravities are excluded by the relative strengths of Balmer lines, and the
strength of the Balmer jump seen in the medium-resolution spectrum. Hence we
conclude that the star is located on the red giant branch. By interpolation in
a 12-Gyr pre-helium-flash stellar evolutionary track$^{9}$, we estimate the
surface gravity of {\he} to be $\log(g) = 2.2 \pm 0.3$, and its mass $M
\approx 0.8\,M_{\odot}$. We derive $\mbox{[Fe/H]} = -5.3 \pm 0.2$ for {\he},
where $\mbox{[A/X]} = \log_{10} (N_{\mbox{\scriptsize A}}/N_{\mbox{\scriptsize
    X}}) - \log_{10}(N_{\mbox{\scriptsize A}}/N_{\mbox{\scriptsize
    X}})_{\odot}$, and the subscript '$\odot$' refers to the Sun. In the
determination of the iron abundance, as well as the Fe\,I/Fe\,II ionisation
equilibrium, we took into account deviations from local thermodynamical
equilibrium (LTE) in the formation of iron lines in the atmosphere of the
star, which led to a correction of $+0.1$ dex for the Fe\,I abundance. The
quoted error in the iron abundance includes uncertainties in the derived model
atmosphere parameters {\tefft} (resulting in $\delta\mbox{[Fe/H]} = \pm
0.20$\,dex) and $\log(g)$ ($\delta\mbox{[Fe/H]} = \pm 0.02$\,dex), the adopted
oscillator strengths ($\delta\mbox{[Fe/H]} = \pm 0.1$\,dex), and in the
line-strength measurement uncertainties ($\delta\mbox{[Fe/H]} = \pm
0.07$\,dex).  An LTE analysis, conducted differentially with respect to {\cd}
(with $\mbox{[Fe/H]} = -3.98$, previously the most iron-deficient giant star
known)$^{1}$, shows that {\he} is 1.4\,dex more iron-poor.  This is in very
good agreement with our non-differential LTE value of $\mbox{[Fe/H]} = -5.4$
for {\he}.

Extreme iron deficiencies have also been observed in some post-asymptotic
giant branch (post-AGB) stars$^{10,11}$.  However, extensive studies have
revealed that their observed elemental abundances were altered substantially
from their primordial compositions by selective dust depletion and subsequent
radiation-driven gas-dust separation$^{10,12}$. The most extreme cases are
HR~4049 and HD~52961, which have $\mbox{[Fe/H]} = -4.8$. Other elements, such
as Ca and Mg, are depleted by a similar amount, while the abundances of
elements such as C, N, O, and Zn are close to the solar values$^{11}$. For
{\he}, on the other hand, we derive an upper limit for the Zn abundance that
is significantly below solar ($\mbox{[Zn/H]} < -2.7$). We also note that these
post-AGB stars occupy a stellar physical parameter space$^{11}$ that is not
shared by {\he}, that is, $\teffm = 6,000$--$7,600$\,K, $\log(g) = 0.5$--$1.2$.
Furthermore, typically many lines in the spectra of post-AGB stars exhibit
prominent emission features owing to a strong stellar wind, while other lines have
a complex absorption structure resulting from the presence of circumstellar
gas$^{11,13}$. Such features are not seen in the spectrum of {\he}. Finally, we
note that infrared photometry of our star does not reveal any indications of
an excess of infrared flux due to hot dust.

We conclude that {\he} likely formed from a gas cloud with a metal abundance
corresponding to $\mbox{[Fe/H]} \approx -5.3$. The abundance pattern of
elements heavier than Mg can be well fit by the predicted elemental
yields$^{14}$ of a $20$--$25\,M_{\odot}$ star that underwent a type II
supernova explosion, indicating that the gas cloud from which {\he} formed
could have been enriched by such a supernova. (See Table 1.) Alternatively,
{\he} could have formed from a zero metallicity gas cloud, with its present
metallicity being due to accretion of material during repeated passages
through the Galactic disc$^{15}$.

The large overabundances of C and N, and possibly Na, in {\he} can be
explained by either mass transfer from a previously more massive companion
during its AGB phase, or else by self-enrichment. The mass-transfer scenario
has been proposed as the likely explanation for the so-called metal-poor CH
stars$^{16}$. Recent model computations of the structure, evolution and
nucleosynthesis of low-mass and intermediate-mass stars of zero or near-zero
metallicity have shown$^{17-19}$ that such stars undergo extensive mixing
episodes at or shortly after the helium-core flash, resulting in a dredge-up
of helium burning and CNO-cycle processed material to their surfaces,
especially carbon and nitrogen. The surface abundance ratios of CNO, Na, and
Mg, predicted by Siess \emph{et al.}$^{19}$ for stars in the mass range\\
$1$--$2\,M_{\odot}$, agree reasonably well with the abundances observed in
{\he}, within the large uncertainties arising from the input physics adopted
in the model calculations.

Our derived upper limits for Ba and Sr indicate that {\he} is not strongly
overabundant in neutron-capture elements. In this respect, our star is similar
to the extremely metal-poor giant$^{20}$ CS~22957$-$27, a star with
$\mbox{[Fe/H]} = -3.4$ and $\mbox{[C/Fe]} = +2.2$, as well as to the recently
discovered class of mildly carbon-enhanced metal-poor stars that are
underabundant in neutron-capture elements$^{21}$. This abundance pattern
implies that helium burning and some CNO burning may have occurred. However,
if this scenario is correct, the observations suggest that the reactions
involved have produced litttle $^{13}$C, or that the $^{13}$C produced has
been consumed by proton captures. We note that if $\alpha$ captures on
$^{13}$C were dominant, neutrons would have been produced through the reaction
$^{13}\mbox{C}(\alpha,\mbox{n})^{16}$O, giving rise to s-process
nucleosynthesis.

\begin{table}[htb]
  \begin{center}
    \begin{tabular}{lr}\hline\hline
      Element & [X/Fe]\rule{0.0ex}{2.3ex}\\\hline
      Li & \rule{0.0ex}{2.3ex} $< 5.3$\\
      C  &  $  4.0$\\
      N  &  $  2.3$\\
      Na &  $  0.8$\\
      Mg &  $  0.2$\\
      Ca &  $  0.4$\\
      Ti &  $ -0.4$\\
      Ni &  $ -0.4$\\
      Zn &  $ <2.7$\\
      Sr &  $<-0.5$\\
      Ba &  $< 0.8$\\
      Eu &  $< 2.8$\\\hline\hline
    \end{tabular}
    \caption{\footnotesize Abundance ratios [X/Fe] of {\he} as derived from a
      high-resolution, high-S/N UVES spectrum. In our analysis, we used a
      custom plane-parallel model atmosphere with the most recent atomic and
      molecular opacity data.  Typical errors in the logarithmic abundances,
      resulting from uncertainties in the stellar parameters and oscillator
      strengths, are $0.1$--$0.2$\,dex. Possible systematic errors are judged
      to be of the same order of magnitude. The abundances of C, N, and Ca
      have been derived from spectrum synthesis, using the C$_2$ band at
      $\lambda = 516.5$\,nm, the CN band at $\lambda = 388.3$\,nm (assuming a
      C abundance as listed above), and the Ca~II~H$+$K lines, respectively.
      We measure a carbon isotopic ratio of $^{12}\mbox{C}/^{13}\mbox{C} > 30$
      from CH A-X lines.}
  \end{center}
\end{table}

It was once believed that the lowest-metallicity objects to have formed in the
Galaxy (at least those that survived until the present) were the halo globular
clusters, such as M92, with $\mbox{[Fe/H]} = -2.5$, a factor of 300 times
below the solar value.  This belief was based on several assumptions, most
importantly that star formation in the early Galaxy would have been strongly
inhibited at low masses, owing to the difficulty of forming stars from nearly
primordial gas without cooling channels arising from heavy elements such as
iron. A number of early studies$^{22,23}$ suggested that cooling from
molecular species including hydrogen might allow low-mass star formation
before the production of heavy metals significantly polluted the interstellar
medium, but no examples of stars approaching such low metallicities as {\he}
were then known.  This view received support from early objective-prism
surveys$^{3}$ that failed to detect significant numbers of low-mass stars with
$\mbox{[Fe/H]} < -2.5$. For the past two decades, more extensive surveys have
pushed the low-metallicity limit to $\mbox{[Fe/H]} = -4.0$, that is, the iron
abundance of {\cd}, but no lower$^{4}$. On the basis of the numbers of
metal-poor stars thus far identified in these surveys, which include some
$1,000$ stars with $\mbox{[Fe/H]} < -2.0$, a simple extrapolation of the
distribution of metal abundances suggested that if stars with $\mbox{[Fe/H]} <
-4.0$ existed in the Galaxy, at least a handful should have been found. The
dwarf carbon star G~77$-$61 has been suggested to have$^{24}$ $\mbox{[Fe/H]} =
-5.5$, if a logarithmic solar Fe abundance of $7.51$ (on a scale where the
logarithmic abundance of hydrogen is $12$) is adopted, as we did for our
analysis of {\he}. However, the analysis of the cool ($\teffm =
4,200$\,K)$^{24}$ dwarf G~77$-$61 is much less certain than the analysis of
{\he}, because the spectrum of the former is dominated by molecular bands of
carbon that are much stronger than in {\he}.  Clearly, the existence of at
least one example of a star with an iron abundance as low as $\mbox{[Fe/H]} =
-5.3$ provides evidence that the limiting metallicity of halo stars may not
yet have been reached.

This has implications for the nature of the first mass function (FMF) of early
star formation. Although some studies$^{5,6}$ have suggested that metal-poor
gas clouds fragment into low-mass objects only above a certain critical
metallicity, $Z_{\mbox{\scriptsize crit}} \approx 10^{-4}\,Z_{\odot}$
($\mbox{[Fe/H]} = -4$), our discovery provides evidence in favor of other
studies suggesting that the FMF included not only very massive stars but also
lower-mass stars$^{25}$. The FMF may also have been bimodal$^{26}$ and may
have included stars with masses around $1\,M_{\odot}$ as well as around
$100\,M_{\odot}$. Other authors$^{27}$ have speculated on the possible
presence of ancient (presumably extremely low metallicity) objects that could
have formed before re-ionization of the early Universe. In this view, the
`gap' between the iron abundance of {\he} and other extremely metal-poor stars
may represent the period of reheating and re-ionization, when star formation
was strongly suppressed owing to destruction of the molecular hydrogen that
was needed for cooling.  As an alternative, it remains possible that the
$\mbox{[Fe/H]} = -4.0$ `limit' is an artefact caused by the brighter magnitude
limits of surveys that have been carried out up to now.  Indeed, it may be no
coincidence that the fainter magnitude limit of the HES allowed for the
detection of {\he} at a distance of about 11\,kpc from the Sun, whereas most
previous surveys have been limited to searches for extremely metal-poor stars
within the inner halo of the Galaxy. Further tests of this hypothesis are
currently targeting the faintest giants from the HES, which extend to 20 or
more kpc from the Sun. If additional stars with iron abundances substantially
lower than $\mbox{[Fe/H]} = -4.0$ are identified, these will provide important
new tools with which, for example, we could directly compare observed
elemental abundance patterns with the predicted yields of the first generation
of supernova type~II explosions, make a definitive measurement of the primordial
lithium abundance (which strongly constrains `big bang' nucleosynthesis)$^{28}$,
and obtain tighter constraints on the nature of the FMF. If any examples of
the so-called r-process-enhanced stars are found with $\mbox{[Fe/H]} < -4.0$,
we may be able to use nucleochronometry$^{29}$ to obtain a direct estimate of
the epoch of first star formation in the Universe, possibly resulting in an
improved lower limit for the age of the Universe.

{\footnotesize
  \begin{flushleft}
    Received 1 July; accepted 24 September 2002\\
    \rule{7.7cm}{0.02cm}
  \end{flushleft}

\begin{enumerate}
\item[1.] Norris, J.E., Ryan, S.G. \& Beers, T.C. Extremely metal-poor stars.
  VIII. High-resolution, high signal-to-noise ratio analysis of five stars
  with $\mbox{[Fe/H]}<-3.5.$ \emph{Astrophys. J.} \textbf{561}, 1034--1059
  (2001)
\item[2.] Cohen, J.G., Christlieb, N., Beers, T.C., Gratton, R. \& Carretta,
  E. Stellar Archaeology: A Keck pilot program on extremely metal-poor stars
  from the Hamburg/ESO survey. I. Stellar parameters. \emph{Astron. J.}
  \textbf{124}, 470--480 (2002)
\item[3.] Bond, H.E. Where is population III? \emph{Astrophys. J.}
  \textbf{248}, 606--611 (1981)
\item[4.] Beers, T.C. In: \emph{The Third Stromlo Symposium: The Galactic
    Halo} (eds. Gibson, B.K., Axelrod, T.S. \& Putman, M.E.) \emph{Astron.
    Soc. Pacif. Conf. Ser.} \textbf{165}, 202--212 (1999)
\item[5.] Bromm, V., Ferrara, A., Coppi, P.S. \& Larson, R. B. The
  fragmentation of pre-enriched primordial objects. \emph{Mon. Not. R. Astron.
    Soc.} \textbf{328}, 969--976 (2001)
\item[6.] Schneider, R., Ferrara, A., Natarajan, P. \& Omukai, K. First stars,
  very massive black holes, and metals. \emph{Astrophys. J.} \textbf{571},
  30--39 (2002)
\item[7.] Wisotzki, L. \emph{et al.} The Hamburg/ESO survey for bright QSOs.
  III. A large flux-limited sample of QSOs. \emph{Astron. Astrophys.}
  \textbf{358}, 77--87 (2000)
\item[8.] Christlieb, N. \emph{et al.} The stellar content of the Hamburg/ESO
  survey.  I. Automated selection of DA white dwarfs. \emph{Astron.
    Astrophys.} \textbf{366}, 898--912 (2001)
\item[9.] Yi, S. \emph{et al.} Towards better age estimates for stellar
  populations: The Y2 isochrones for solar mixture. \emph{Astrophys. J.
    Suppl.} \textbf{136}, 417--437 (2001)
\item[10.] Mathis, J.S. \& Lamers, H.J.G.L.M. The origin of the extremely
  metal-poor post-AGB stars. \emph{Astron. Astrophys.} \textbf{259}, L39--L42
  (1992)
\item[11.] Van Winckel, H., Waelkens, C. \& Waters, L.B.F.M. The extremely
  iron-deficient ``post-AGB'' stars and binaries. \emph{Astron. Astrophys.}
  \textbf{293}, L25--L28 (1995)
\item[12.] Waters, L.B.F.M., Trams, N.R. \& Waelkens, C. A scenario for the
  selective depletion of stellar atmospheres. \emph{Astron. Astrophys.}
  \textbf{262}, L37--L40 (1992)
\item[13.] Bakker, E.J. \emph{et al.} The optical spectrum of HR 4049.
  \emph{Astron.  Astrophys.} \textbf{306}, 924--934 (1996)
\item[14.] Woosley, S. E. \& Weaver, T.A. The evolution and explosion of
  massive stars. II. Explosive hydrodynamics and nucleosynthesis.
  \emph{Astrophys. J.} \textbf{101}, 181--235 (1995)
\item[15.] Yoshii, Y. Metal enrichment in the atmospheres of extremely
  metal-deficient dwarf stars by accretion of insterstellar matter.
  \emph{Astron. Astrophys.} \textbf{97}, 280--290 (1981)
\item[16.] McClure, R.D \& Woodsworth, A.W. The binary nature of the Barium
  and CH stars. III. Orbital parameters. \emph{Astrophys. J.} \textbf{352},
  709--723 (1990)
\item[17.] Fujimoto, M.Y., Ikeda, Y. \& Iben Jr., I. The origin of extremely
  metal-poor carbon stars and the search for population III. \emph{Astrophys.
    J.}  \textbf{529}, L25--L28 (2000)
\item[18.] Schlattl, H., Salaris, M., Cassisi, S. \& Weiss, A. The surface
  carbon and nitrogen abundances in models of ultra metal-poor stars.
  \emph{Astron.  Astrophys.} (2002) (in the press); also preprint
  astro-ph/0205326 at http://xxx.lanl.gov
\item[19.] Siess, L., Livio, M. \& Lattanzio, J. Structure, evolution, and
  nucleosynthesis of primordial stars. \emph{Astrophys. J.} \textbf{570},
  329--343 (2002)
\item[20.] Norris, J., Ryan, S.G. \& Beers, T.C.  Extremely metal poor stars.
  The carbon-rich, neutron capture element-poor object CS 22957--027.
  \emph{Astrophys. J.} \textbf{489}, L169--L172 (1997)
\item[21.] Aoki, W., Norris, J.E., Ryan, S.G., Beers, T.C. \& Ando, H. The
  chemical composition of carbon-rich, very metal poor stars: a new class of
  mildly carbon rich objects without excess of neutron-capture elements.
  \emph{Astrophys. J.} \textbf{567}, 1166--1182 (2002)
\item[22.] Yoshii, Y. \& Sabano, Y. Stability of a collapsing pre-galactic gas
  cloud. \emph{Publ. Astron. Soc. Jpn.} \textbf{31}, 505--521 (1979)
\item[23.] Palla, F., Salpeter, E.E. \& Stahler, S.W. Primordial star
  formation: the role of molecular hydrogen. \emph{Astrophys. J.}
  \textbf{271}, 632--641 (1983)
\item[24.] Gass, H, Liebert, J. \& Wehrse, R. Spectrum analysis of the
  extremely metal-poor carbon dwarf star G77$-$61. \emph{Astron. Astrophys.}
  \textbf{189}, 194--198 (1988)
\item[25.] Yoshii, Y. \& Saio, H. Initial mass function for zero-metal stars.
  \emph{Astrophys. J.} \textbf{301}, 587--600 (1986)
\item[26.] Nakamura, F. \& Umemura, M. On the initial mass function of
  population III stars. \emph{Astrophys. J.} \textbf{548}, 19--32 (2001)
\item[27.] Ostriker, J.P. \& Gnedin, N.Y. Reheating of the universe and
  population III. \emph{Astrophys. J.} \textbf{472}, L63--L67 (1996)
\item[28.] Ryan, S.G., Norris, J.E. \& Beers, T.C. The Spite lithium plateu:
  ultrathin but postprimordial. \emph{Astrophys. J.} \textbf{523}, 654--677
  (1999)
\item[29.] Cayrel, R. \emph{et al.} Measurement of stellar age from uranium
  decay.  \emph{Nature} \textbf{409}, 691--692 (2001)
\end{enumerate}

\noindent {\bf Acknowledgements} We thank the European Southern Observatory for
providing us with reduced UVES spectra. We are grateful to M. Asplund, B.
Edvardsson, J. Lattanzio, J. Norris, N. Piskunov, B. Plez, D. Reimers, S.G.
Ryan, L. Siess and L. Wisotzki for their contributions and suggestions. N.C.
acknowledges a Marie Curie Fellowship granted by the European Commission, and
support from Deutsche Forschungsgemeinschaft. T.C.B. acknowledges grants of
the US National Science Foundation, S.R. support from FAPESP and CNPq, and the
Uppsala group from the Swedish Research Council.\\

\noindent {\bf Competing interests statement} The authors declare that they have no
competing financial interests.\\

\noindent {\bf Correspondence} and requests for materials should be addressed to N.C.
(e-mail: nchristlieb@hs.uni-hamburg.de).
}

\end{document}